\documentclass[]{iopart}
\usepackage{latexsym}
\usepackage{graphicx}
\usepackage[dvips,usenames]{color}
\usepackage{epsfig}

\begin{document}

%%%%%%%%%%%%% style macros

\newcommand{\va}{\vspace{5mm}}
\newcommand{\vb}{\vspace{10mm}}
\newcommand{\vs}{\vspace{15mm}}

\newcommand{\tc}[2]{\textcolor{#1}{#2}}       % 1 fg color, 2 text
\newcommand{\cb}[2]{\colorbox{#1}{#2}}        % 1 bg color, 2 text
\newcommand{\fb}[3]{\fcolorbox{#1}{#2}{#3}}   % 1 frame, 2 bg, 3 text

% \color{#1} changes fg color to named color #1
% \color[rgb]{#1,#2,#3} changes fg color to rgb color=(1,2,3).

\parindent0mm
\fboxsep3mm
\fboxrule1mm

%%%%%%%  equation macros

\newcommand{\be}{\begin{equation}}
\newcommand{\ee}{\end{equation}}
\newcommand{\ba}{\begin{eqnarray}}
\newcommand{\ea}{\end{eqnarray}}
\renewcommand{\NL}{\nonumber \\}

\newcommand{\binom}[2]{ \left( \begin{array}{c} #1 \\ #2 \end{array} \right)}
\newcommand{\E}[1]{ \langle #1 \rangle }
\newcommand{\ov}[1]{ \overline{#1} }

%%%%%%%%% FIG macro %%%%%%%%%%

\newcommand{\inplot}[2]{
 \begin{center}
 \leavevmode\epsfysize=#2mm \epsfbox{#1}
 \end{center}
}

%%%%%%%  special macros

% \newcommand{\PG}{PineGreen}
\newcommand{\PG}{OliveGreen}
\newcommand{\cpg}{\color{\PG}}

%%%%%%%%% header  definition

\def\fejlec{
  \hrulefill \qquad {\bf T.S.Bir\'o: SQM'03 theoretical summary
   } \qquad  \hrulefill \ \
}

%% normal page style uses foil headings

%% \pagestyle{headings}
%% \leftheader{{\color{RoyalPurple} \fejlec}}
%% \rightheader{{\color{RoyalPurple} \fejlec}}

%%%%%%%%%%%  front page

%%%%%%%%%%%%%%%%%%%%%%%% TITLE PAGE %%%%%%%%%%%%%%%%%%%

\thispagestyle{empty}

%\begin{center}
%\fb{WildStrawberry}{White}{{\bf \Large  
%  Strange Quark Matter Theory
%}}
% \va
% \fb{WildStrawberry}{White}{{\bf \Large  
% }}

\title[SQM2003 theoretical summary]{Strange Quark Matter Theory}

%\vspace{10mm}

%{\sc \tc{BlueViolet}{T.S. Bir\'o}\footnote{
% Theoretical summary talk at SQM03, March 12-17, Atlantic Beach, NC, USA.
% }
%}

\author{Tam\'as S. Bir\'o}
\address{ KFKI Res.Inst.Part.Nucl.Physics, 
          H-1525 Budapest Pf. 49, Hungary}
\ead{tsbiro@sunserv.kfki.hu}

%\end{center}

%\centerline{\small MTA RMKI, H-1525 Budapest, Pf. 49, Hungary}

% \vspace{5mm}
% \centerline{Version \tc{Fuchsia}{1.1.0}}

%\maketitle

%%%%%%%%%%% main text body

%%%%%%%%%%%%%%%%%%%%%%%%%%%% PAGE 0 %%%%%%%%%%%%%%%%%%%%%%%%%%%
%\newpage

\vspace*{4mm}
Facing the tantalizing amount of experimental and simulated or
computational data which were shown us at this meeting I was
tempted to say that now the task of theorists would be to
describe (these) data. Then I started thinking. And realized: no!
I am objecting this view. Our task is not to {\em describe},
but to {\em understand} data.

\va
This effort to understand, whereas understanding necessarily
involves reduction (not throwing, however, really important parts away),
implies further efforts. Here is a to-do list of the necessary 
prerequisites of understanding:
\begin{itemize}
\item	{understand theory} (this might sound trivial first,
	but I assure it is not),
\item	{verify assumptions} (assumptions are always there,
	explicit or implicit, when applying abstract theoretical
	concepts in models of reality)
\item	{falsify assumptions} (equally important as verification,
	otherwise different theories would interpret the same data)
\item	{weave a network of non-contradicting statements }
	(which is still a process internal to the theorist's work), and
	finally
\item	{make a new theory only if it is unavoidable}
	(otherwise we would be led too much by our phantasy and
	mathematical invention instead of experimental facts).

\end{itemize}

\va
Ladies and gentlemen, this is going to be the theoretical summary talk
of the Strange Quark Matter 2003 conference. When I was alerted by
the e-mail we all got, ``prepare your transparencies'',
I took this home-work exercise seriously. I have prepared quite a few
pages {\em before} this conference. What can one know in advance,
before listening to the talks?\footnote{I attended to almost all of them,
 and I listened to all I attended}.

\va
First of all there is a general outline which a summary talk should
follow. On the level of the basic theory one is supposed to conclude
about the present status of the underlying theoretical concepts,
one ought to emphasize important news, the novel aspects we are
encountering, and finally it is useful to formulate in a possibly
definite way, what our perspectives for further development are.

\va
A summary of the phenomenology should also be given, not necessarily
in the order of their original presentation, but restructured,
following the natural clusters and network of themes and applied
methods which is formed by the contemporary research.
Alas, this dichotomic splitting to basic theory and phenomenology
is {\em basically} false: reality is rather described by a
continuous spectrum than by two peaks, there are even cases of
mixed states.

\va
A separate effort should be devoted to selecting out 
{highlights}, 
{new predictions and explanations}, if any occurred.
Some colleagues used to apply the defaming expression ``post-dictions''
to explanations. This attitude -- I think -- is mistaken.
Really good explanations of already known experimental facts
help us towards {\em understanding}, especially when, 
if they are selective, clearly separating alternative approaches.
The key requirement is to contribute to our understanding.

\va
Finally, in order to complete the summary, I have written up a list
of issues {``we did not talk about''}, meaning to mention  
theoretical works and concepts not taken explicitly into the
official program, although implicitly used and often referred to
by speakers in their presentations.

\va
After attending to the lectures, listening to the talks and 
a seemingly uncountable number of follow-up discussions I was able
to formulate a second summary outline. This I would like to
present in a form of questions. The main question of this and
several previous meetings remained:

\va
\centerline{{Have we seen (strange) quark matter?}}

\va
This question implies further questions.
What is actually quark matter? Is there any quark-level collectivity?
The concept of collectivity can be split to the following main
subtopics:
\begin{itemize}
\item	the question of {\bf flow and $T$} (meaning a common slope of
	transverse momentum spectra of different particle species,
	not temperature),
\item	quark level {\bf recombination} and its effects on the
	finally observed hadronic states,
\item	{\bf suppression} and/or {\bf enhancement} of any
	hadronic or leptonic signal known from $pp$ and $e^+e^-$
	collisions, and finally
\item	the very question whether all what we see is just a
	{\bf maximum -- entropy state}, washing out all memory of
	that, whatever may have existed before.
\end{itemize}
Can we trace back the pre-hadron state? Do we understand already
what quark matter is or are we misled by preconception and
wishful thinking of simplicity by doing the theorists' homework?

\va
\begin{figure}
\begin{center}
\includegraphics[width=0.5\textwidth,height=0.4\textwidth,bb=38 69 558 772]{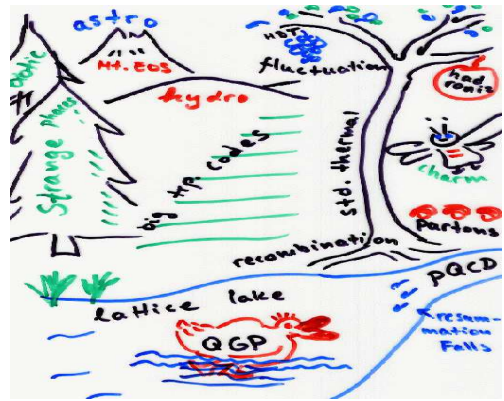}
\end{center}
\caption{\label{Fig1} An artists' view of the theory landscape at
the Strange Quark Matter 2003 Conference.}
\end{figure}

\va
In order to present an overview of the theoretical topics of this
conference I would like to confront you with an ``artist's view''
of the theory landscape of SQM 2003 (Fig.\ref{Fig1}).
This landscape includes a lake with a big, fat duck swimming in it
at the front, a small stream rushing into the lake, a few big trees,
green fields in the background and far-away hills and mountains.
I have learned from the talk given by Barbara Jacak -- and I think
she was citing Nu Xu -- that the quark gluon plasma (QGP)
{``looks like a duck and talks like a duck''}
(the experimentalists just cannot say whether what they see is a duck).
The duck first has been spotted in the {Lattice Lake} which
receives some fresh water from {pQCD} 
(perturbative QCD) stream
through the Resummation Falls.

\va
There is a tree behind with ramified roots near to the water,
symbolizing the {parton model} and in particular 
the quark level recombination. On the other hand a unique, plain-textured
stem dives high from this, the {``standard thermal model''}.
Its main and only advantage lies in simplicity. It may dart some shadow
onto the duck, but can hardly help to learn what the duck looks like.
The branches of the tree carry uncountable leaves representing
{fluctuation and correlation} studies, and a big,
reddish fruit is starting to reach the state of maturity, the
apple of {hadronization}.

\va
The big vast field in the background stands for the big
{micro-dynamical codes}. At a distance they all show
the same green color, only by laying down can one start to see
separate grass pieces. It requires close inspection. Beyond all
those smooth hills of {hydro} and the Mt.EoS,
symbolizing the far leading problem of the
{equation of state}, occur. The top of this mount,
where all climbers must be heading to, may -- by the way --
be reflected in the water of the Lattice Lake quite accurately.
Above, mostly in cloudy skies, reside 
{astrophysics and cosmology}. Finally to our left some
strange looking pine trees hint to a forest of 
{exotic theories}.

\va
A further prepared transparency shows you a clustering of the
theoretical themes presented at this conference, deciphered from
the preliminary program. Up to a few name changes it reflects
now the real program well. The themes in a rather reverse order
of their presentation, were:
lattice QCD, (Steven Gottlieb, P\'eter Petreczky, Owe Philipsen)
{astrophysics,} 
(Mark Alford, Fred Walter, Markus Thoma, Madappa Prakash, James Lattimer)
{strange and exotic,} 
(Johann Rafelski, Volker Koch, Jack Sandweiss, Ariel Zhintnitsky, 
Jeffrey Bowers, Sarmista Banik, J\"urgen Schaffner-Bielich)
{partons, pQCD, charm,} 
(Klaus Werner, Thomas Mehen,Ralf Rapp, Joe Kapusta, Bob Thews, 
Kevin Haglin)
{hadronization and fluctuation,} 
(Rainer Fries, D\'enes Moln\'ar, Chiho Nonaka, Ulrich Heinz, Sangyong Jeon,
Ziwei Lin, Sean Gavin, Laura Tolos, Gary Westfall)
{HBT, correlations,}
(Roy Lacey, Zhangbu Xu, Fabrice Retiere, Huang Huang)
{big micro-dynamical codes,} 
(Marcus Bleicher, Sven Soff, Julia Velkovska, Christoph Hartnack, 
Barbara Jacak)
and {statistical models} 
(Johanna Stachel, Detlev Zschiesche, Fuming Liu, Jean Cleymans, 
George Torrieri).

\va
Now let me reconsider -- in the original time order -- what we have
learned, what I, as a test-person, have learned from the theoretical
talks of this meeting. In the opening review talk given by Ian Rafelski
we were reminded of the basic facts of strange quark matter physics:
\begin{enumerate}
\item	{\em Strangeness is produced}. On the quark level by
	$s\overline{s}$ pair production, on the hadronic level
	by associated hyperon and strange meson production -- so
	actually {\bf net} strangeness is {\bf never} produced
	in strong interaction.
\item	{\em Strangeness is equilibrated}. This fact can be
	really understood only on the quark level, where the
	gluon fusion ($gg \rightarrow s\overline{s}$) cross section
	is large and dominant leading to equilibration in
	fractions of 1 fm/c. On the hadronic level 
	{\em strangeness is not equilibrated}. It is practically 
	impossible due to much longer, in the order of tens of fm/c
	hadrochemical time scales. A caveat to this point: it was
	tacitly assumed that that gluons and gluon fusion is not
	suppressed by some nontrivial, non-perturbative phenomenon.
\end{enumerate}
The central role of {chemistry} 
and the usefulness of thinking in terms
of chemistry was also emphasized by the speaker. We learned that
the equilibrium {chemical potentials} 
($\mu_a$) do not fix all
particle species numbers ($N_i, \quad i > a$). Conventional chemical
potentials are related to conserved charges, to a few selected
linear combinations of particle types ($Q_a = \sum_i q_{a,i} N_i$).
There are other combinations, e.g. those related to pairs, which are
linearly independent from these constraints.
They can be characterized by {fugacities} ($\gamma_b$).
Whenever such a quantity is not one, the system is
{\em not in chemical equilibrium}.

\va
\begin{figure}
\begin{center}
\includegraphics[width=0.4\textwidth,bb=38 407 451 772]{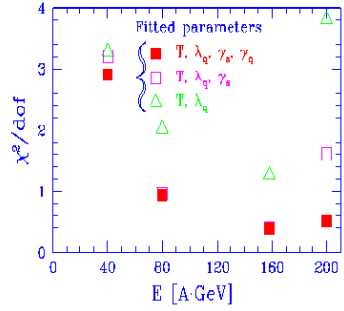}
\end{center}
\caption{\label{Fig2} A fit taking into account fugacities is much
better than without. (From Ian Rafelski's talk.)}
\end{figure}

\va
It was made plain by him with a definite 
{statement}: a $\chi^2/$degree
of freedom fit is much better using $\gamma_{{\rm strange pairs}} < 1$,
than without this possibility taken into account (Fig.\ref{Fig2}).
Later (by others) an equally firm {counterstatement}
was made: $\gamma \approx 0.75$ does not matter (does not quark matter).
Last but not least to this point I would like to emphasize that
contrary to Johanna Stachel's remark fugacity is not a
``fudge factor''. {\em It is as fundamental as the chemical potential
itself}, no less, no more.

\va
\begin{figure}
\begin{center}
\includegraphics[width=0.5\textwidth,angle=-90,bb=38 18 586 772]{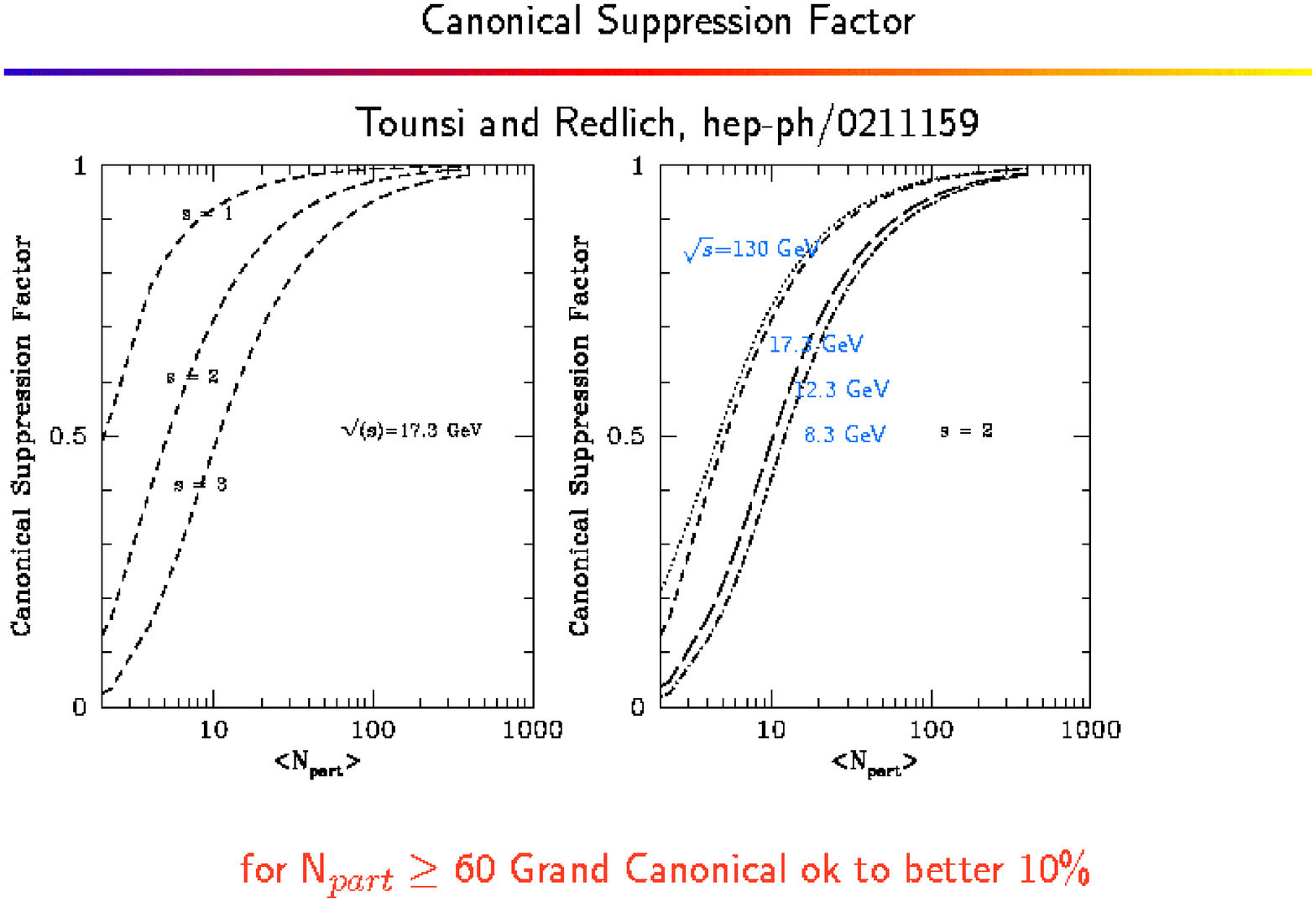}
\end{center}
\caption{\label{Fig3} Strength of the canonical suppression for simply
and multiply strange hyperons. (From Johanna Stachel's talk.)}
\end{figure}

\begin{figure}
%\begin{center}
\includegraphics[width=0.3\textwidth,bb=0 0 595 842]{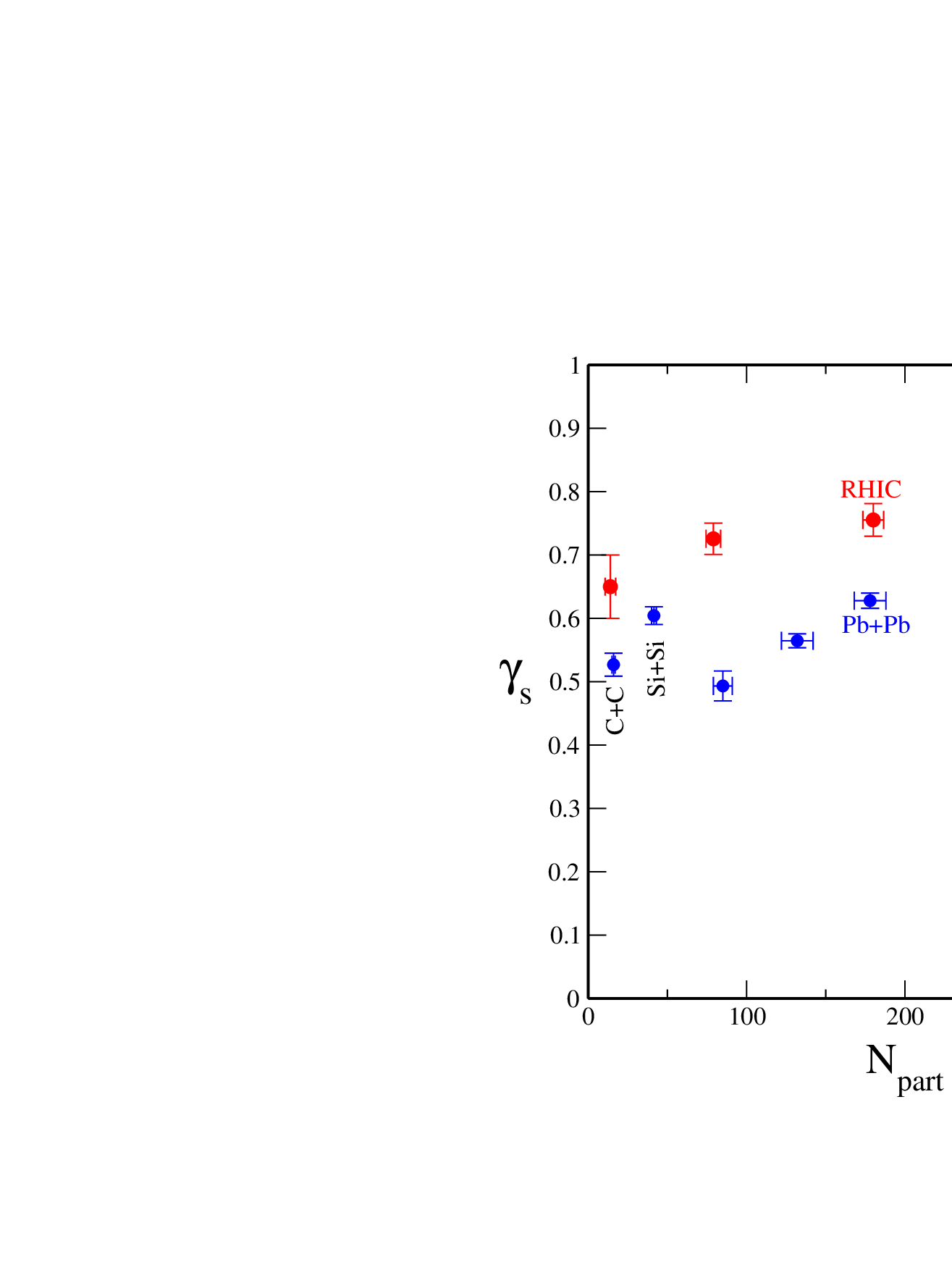}
\includegraphics[width=0.3\textwidth,bb=0 0 595 842]{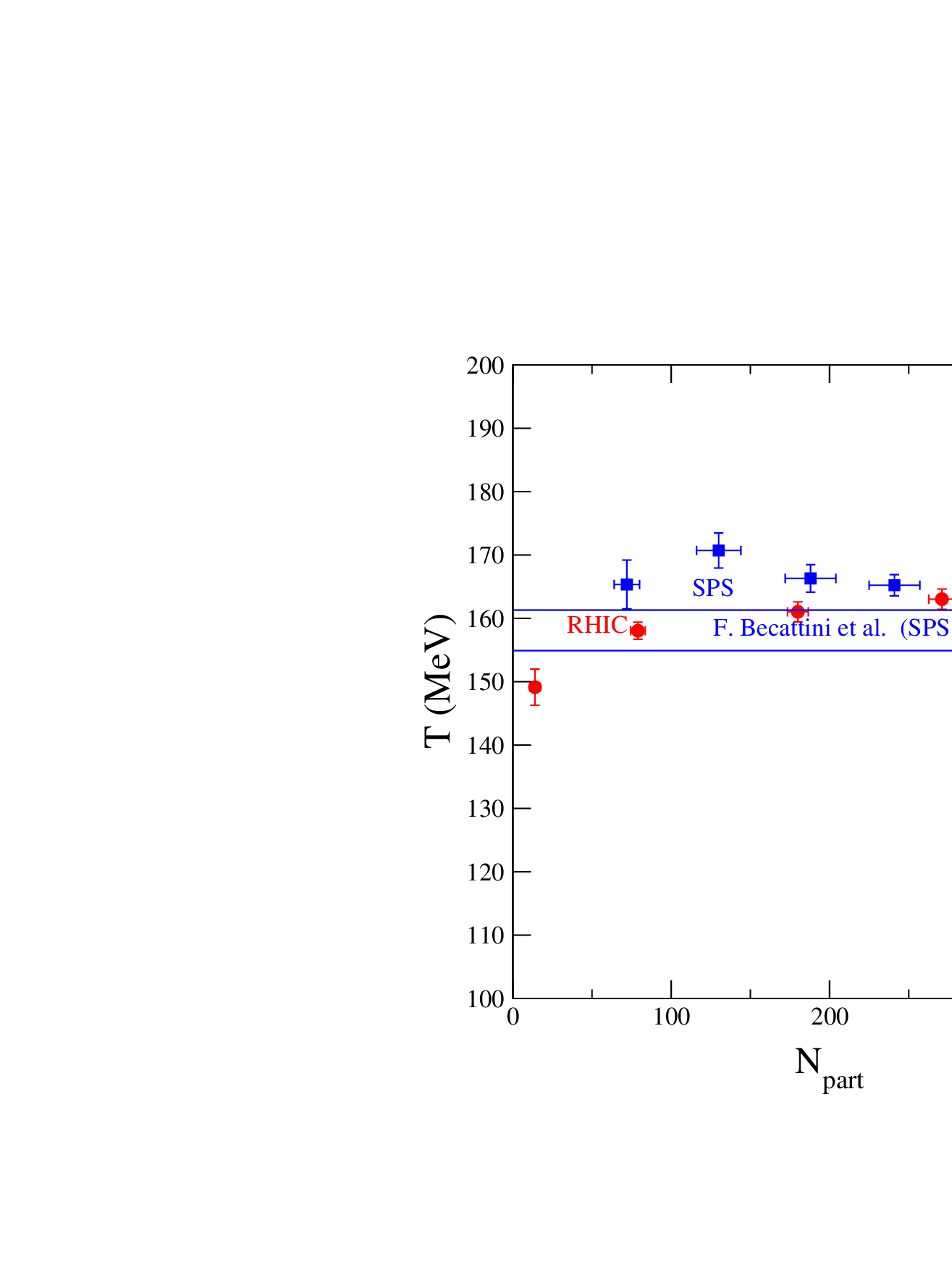}
\includegraphics[width=0.3\textwidth,bb=0 0 595 842]{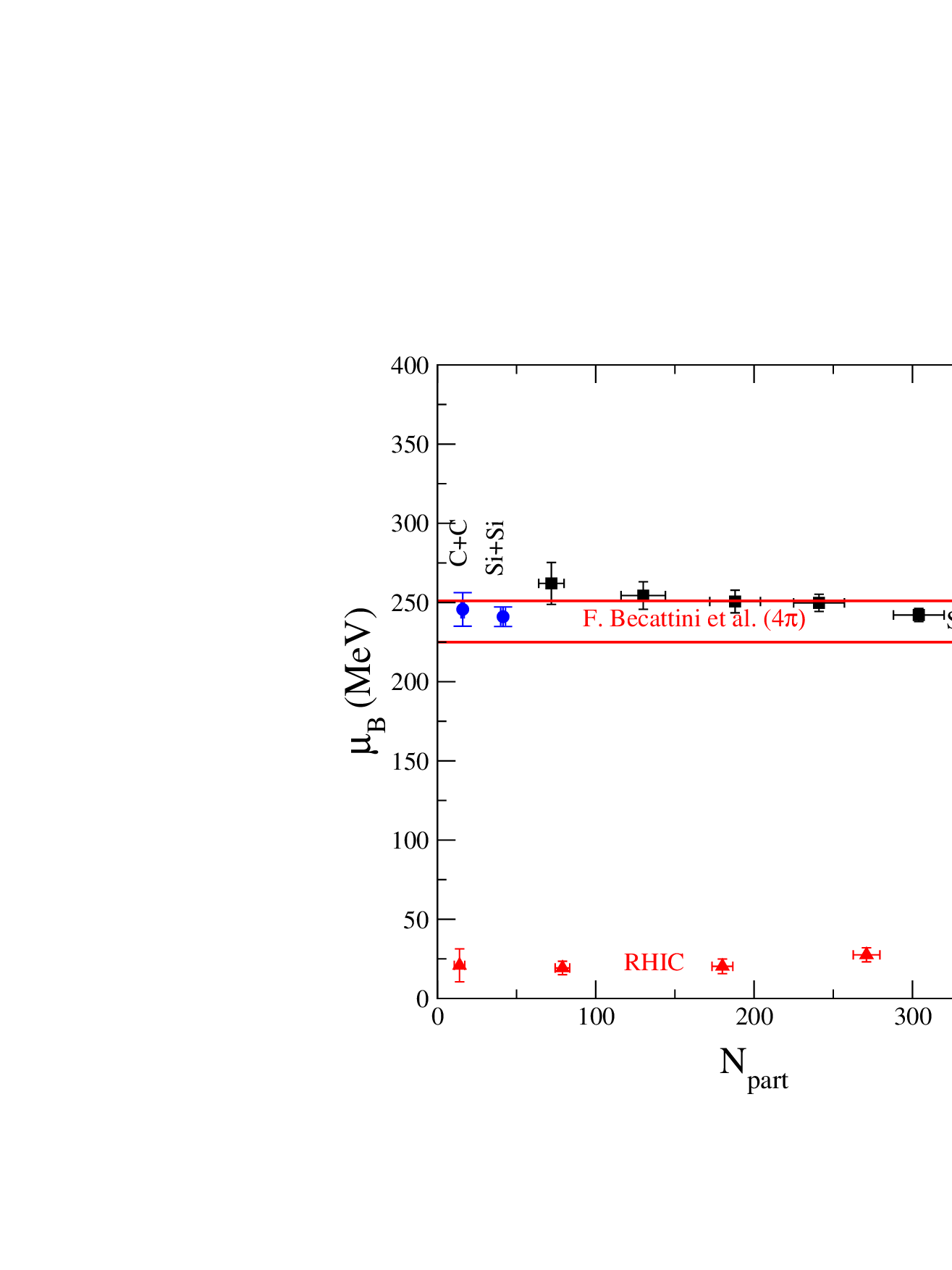}
%\end{center}
\caption{\label{Fig4} Centrality dependence of  fugacities, temperatures
and baryochemical potential in the thermal model. 
(From Jean Cleyman's talk.)}
\end{figure}

\va
We have two principal possibilities of dealing with conservation
laws: i) take them literally and then end up with a
{(micro)canonical} description, or ii) take them on the mean
(hoping that a reservoir is there) and use the
{grand canonical} approach. Where it matters, is the case when
the available phase space is small. This phenomenon was named
-- strangely enough to me -- {``canonical suppression''}
and was promoted by Redlich, Cleymans, Koch, Tounsi and others.
The next figure (Fig.\ref{Fig3}) borrowed from Johanna Stachel's talk shows
how effective this factor really is: from about ten participants on
it becomes unimportant. Jean Cleymans has shown us in his talk
yesterday trends of statistical model parameters as a function of
the participant number. The $\gamma_s$ rises from 0.5 to 0.75 at SPS
and from 0.65 to 0.9 at RHIC, showing a clear trend (Fig.\ref{Fig4}).
The temperature parameter, fitted to spectral slopes, is much less
trendy, it behaves differently at SPS than at RHIC energies.
The most characteristic behavior is shown by the baryon chemical
potential, $\mu_B$: it is namely constant, albeit a different
constant, at both accelerator energies (it shows {\em no
centrality dependence at all}).

\va

\va
Many of us were shocked by the thermal model, more so, that it has
been applied and seems to work not only for heavy ion collisions,
but also for $pp$ and $e^+e^-$ data (Becattini et. al.).
In my opinion we who have been shocked might have been misled
by our gut feelings and have been misjudging the minimal size
which is necessary for the {\em thermal limit} to make sense.
How far can the thermal limit be? This question was also addressed
by Fuming Liu in a micro-canonical analysis of $pp$ data and was
shown that for pions the thermal model works well
(not for heavy hyperons, however).

\va
\begin{figure}
\begin{center}
\includegraphics[width=0.5\textwidth,bb=54 103 576 722]{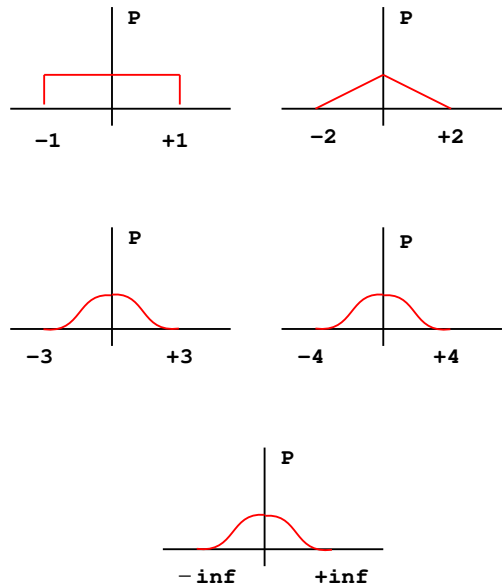}
\end{center}
\caption{\label{Fig5} A sketch of the scaled uniform random 
distribution and the
distribution of the sum of several such random variables.
The infinite number limit is the Gauss distribution.}
\end{figure}

\va
I would like to present here a simple case study which might help
to change our attitude towards what is big and what is small
with respect to thermal models. Take a random variable
$x$ distributed in the interval $(-1,1)$ uniformly (Fig.\ref{Fig5}).
The distribution of the sum of two such variables,
$P_2(x)=\int dx_1 \int dx_2 \delta(x-(x_1+x_2))$, 
shows a triangular shape. The distribution with three
variables is made of parabolic (second order) pieces,
it already shows the main qualitative features (symmetric maximum,
two inflection points) of the Gauss distribution
$$ P_{\infty}(x) = \lim_{n\rightarrow\infty} \, \int \prod_{i=1}^{n}
dx_i \, \delta(x - \sum_{i=1}^n x_i). $$
Already $P_4$ is not easily distinguishable from the
Gaussian (thermal) limit.

\va
Talking about finite systems the main classification is done
according to the value of mean occupancy (number of quanta)
$\langle N \rangle$. If it is much bigger than one, we approach
$\langle N^2 \rangle \approx \langle N \rangle^2$ and everything
looks like a Gaussian distribution. In the opposite case,
when $\langle N \rangle \ll 1$, we have 
$\langle N^2 \rangle \approx \langle N \rangle$ and the generic
distribution is Poissonian. From Volker Koch's review talk
on the topic we could learn that the normalized second cumulative moment
of charge fluctuations, $F_2$, most probably varies in time, so
whenever one is led to infer $F_2 < 1$ from experiments, he/she
would witness a memory of a short time scale process.
Several other talks, e.g. those given by Chiho Nonaka, Sangyong Jeon
and Gary Westfall elaborated on this point. A main message of these
studies is that the {\em finite size scaling} of cumulative
moments is important: equilibrium can be excluded (although not
positively proven) by analysis of scaling powers.

\va
\begin{figure}
\begin{center}
\includegraphics[width=0.5\textwidth,bb=71 427 587 704]{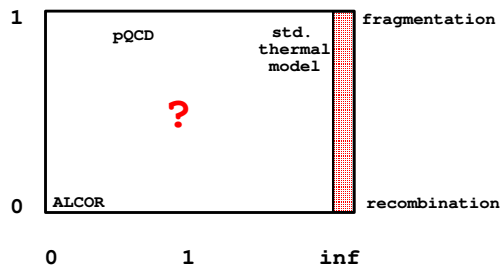}
\end{center}
\caption{\label{Fig6} An abstract map of hadronization models, classified
by the parton fragmentation -- recombination mechanism and the 
characteristic time scale.}
\end{figure}

\va
In general we would like to understand both the
{mechanism} and the {timescale} of the
formation of the observed hadronic final state in high energy heavy
ion collisions. On an abstract map with qualitative scales 
(Fig.\ref{Fig6}),
the horizontal axis of time constants distinguishes between 0, 1 and 
infinity, the vertical axis represents the {\em recombination}
and {\em fragmentation} mechanisms by $0$ and $1$, respectively. 
Here pQCD draws the upper line border of nowadays theoretical approaches,
while micro-dynamical codes cover rather the lower edge.
The effective quark recombination model by Zim\'anyi et.al.
(ALCOR) resides near to the origin, its philosophical counterpart,
the ``standard thermal model'' occupies the right side line at infinite
time scales (assuming equilibrium, so the hadronization mechanism
does not matter in this case). We do not yet know where the real
process of hadronization of a quark-gluon plasma should be
located on this plot, it has a big question mark.

\va
\begin{figure}
\begin{center}
\includegraphics[width=0.5\textwidth,angle=-90,bb=23 40 573 801]{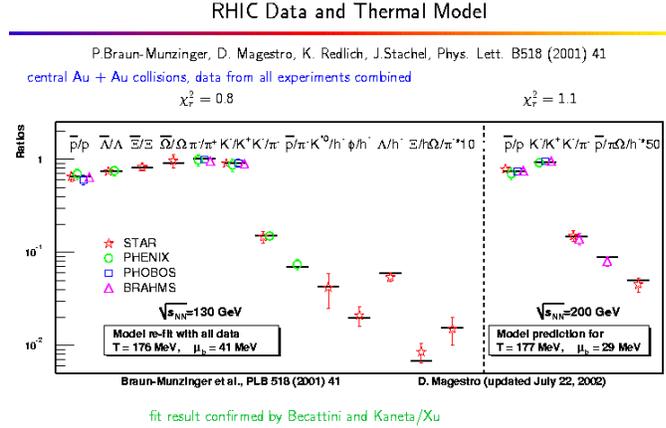}
\end{center}
\caption{\label{Fig7} Logarithmic particle ratios predicted by the
standard thermal model at RHIC energies. (From Johanna Stachel's talk.)}

\end{figure}

\va
As a flavor of success of the thermal model strongly logarithmic
ratios of particle numbers for many selected types were shown at
different accelerator energies by Johanna Stachel. Fig.\ref{Fig7}, plotting
RHIC data is a good representative for the overall quality of
these fits, used to infer bulk parameters such as temperature,
baryon chemical potential, etc. I was very pleased to hear the
speaker's conclusion  at the end of her talk:
{\em ``It works perfectly, but it can't be.''}
I rather agree with the second half of this sentence and I would leave the
word ``perfect'' certainly out from this evaluation,
but otherwise we can witness a novel convergence of opinions here.
That this simple model ``works'' on the other hand imposes on us a real
task to try to understand even the relative success of such
simplifications.

\va
It is clear that we are dealing with highly non-equilibrium
dynamical processes, which lead to a (closely) statistical final state.
{\bf This is not simply ``equilibrium''.} Such a result can occur
via at least three different underlying mechanisms with different
physical background.
\begin{enumerate}
\item	{gain and loss} term rates are nearly equal
	leading to a ``steady state''; this is the textbook case
	for equilibrium (however, this equilibrium is sometimes
	partial only),
\item	{all rates approach zero} fast enough resulting in a
	state in which the pre-history of the mixture is frozen in
	(this process is called ``freeze out''),
\item	a repetitive and {self similar} dynamical process,
	e.g. multiple fragmentation, leads to a pattern reflecting
	self organized criticality (s.o.c.).
\end{enumerate}
Giorgio Torrieri pointed out nicely in his talk where the main
uncertainties of statistical models (which assume a state of maximum
entropy, or thinking of entropy as lack of information, a state of
maximal ignorance) lie. He showed the general formula
$$ E \frac{dN}{d^3p}  \, = \,
\int d\Sigma_{\mu} p^{\mu} \, f(T, p_{\mu}u^{\mu}, \mu_f) \, + \,
{\rm resonances}. $$
The form, evolution and the very nature of a freeze-out hypersurface,
$\Sigma_{\mu}$, may pose restrictions, e.g. causality limits on the
statistical interpretation. This question was copiously studied
in HBT concentrated works on which Roy Lacey, Fabrice Retiere and
Ziwei Lin presented talks. The latter called our attention to
non-trivial $x-t$ correlations which may influence the interpretation
of experimental data.  Furthermore the value of chemical potentials
and the neglect of further fugacity factors is correct only as far
as chemical equilibrium has been established. Finally the presence and
different time evolution of resonances may add new terms to this
simple picture.

\va
\begin{figure}
\begin{center}
\includegraphics[width=0.5\textwidth,bb=35 431 344 740]{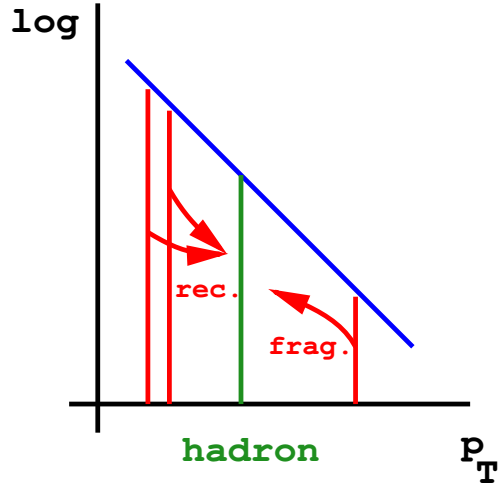}
\end{center}
\caption{\label{Fig8} A sketch of the parton recombination and 
fragmentation mechanisms in the $p_T$ spectra.
(After Rainer Fries' talk.}
\end{figure}

\begin{figure}
\begin{center}
\includegraphics[width=0.5\textwidth,bb=-1 -1 427 342]{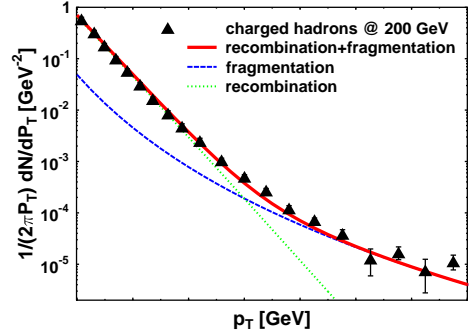}
\end{center}
\caption{\label{Fig9} Transverse momentum spectrum of pions from 
recombination, fragmentation and experimental data.
(From Rainer Fries' talk.}
\end{figure}

\va
These experiences push us back to the roots. The roots of statistical
and bulk models grasp into the {parton dynamics}
which in turn should have its backup from field theory.
Here is quite {\bf new} the application of the idea of quark
{recombination} (Fig.\ref{Fig8}). 
Following Rainer Fries I repeat here the main
idea in its possibly simplest form. Assume that a meson with
transverse momentum $p$ is a recombination of two partons, a quark and
an anti-quark, each with a momentum of $p/2$. With the same logic
baryons are composed of 3 quarks each with a momentum of $p/3$.
$$ {\rm meson}(p) = {\rm quark}^2(p/2), $$
$$ {\rm baryon}(p) = {\rm quark}^3(p/3). $$
Interpreting this as a functional equation it gives us the hint that
the stationary solution of hadronization rate equations is proportional
to $\exp(-p/T)$. On the other hand pQCD calculations using
fragmentation functions always connect a parton with momentum $p$
to hadrons with less momentum $zp (z < 1)$ and -- due to the very
nature of free propagators -- always lead to a power-law behavior
of transverse spectra ($p^{-a}$). Experimental data show a nice interpolation
between these two mechanisms as passing from low to high $p_T$ at
around $4$ GeV/c (Fig.\ref{Fig9}). 
Here we deal with absolutely non-equilibrium
processes.

\va

\begin{figure}
\begin{center}
\includegraphics[width=0.5\textwidth,bb=47 162 549 680]{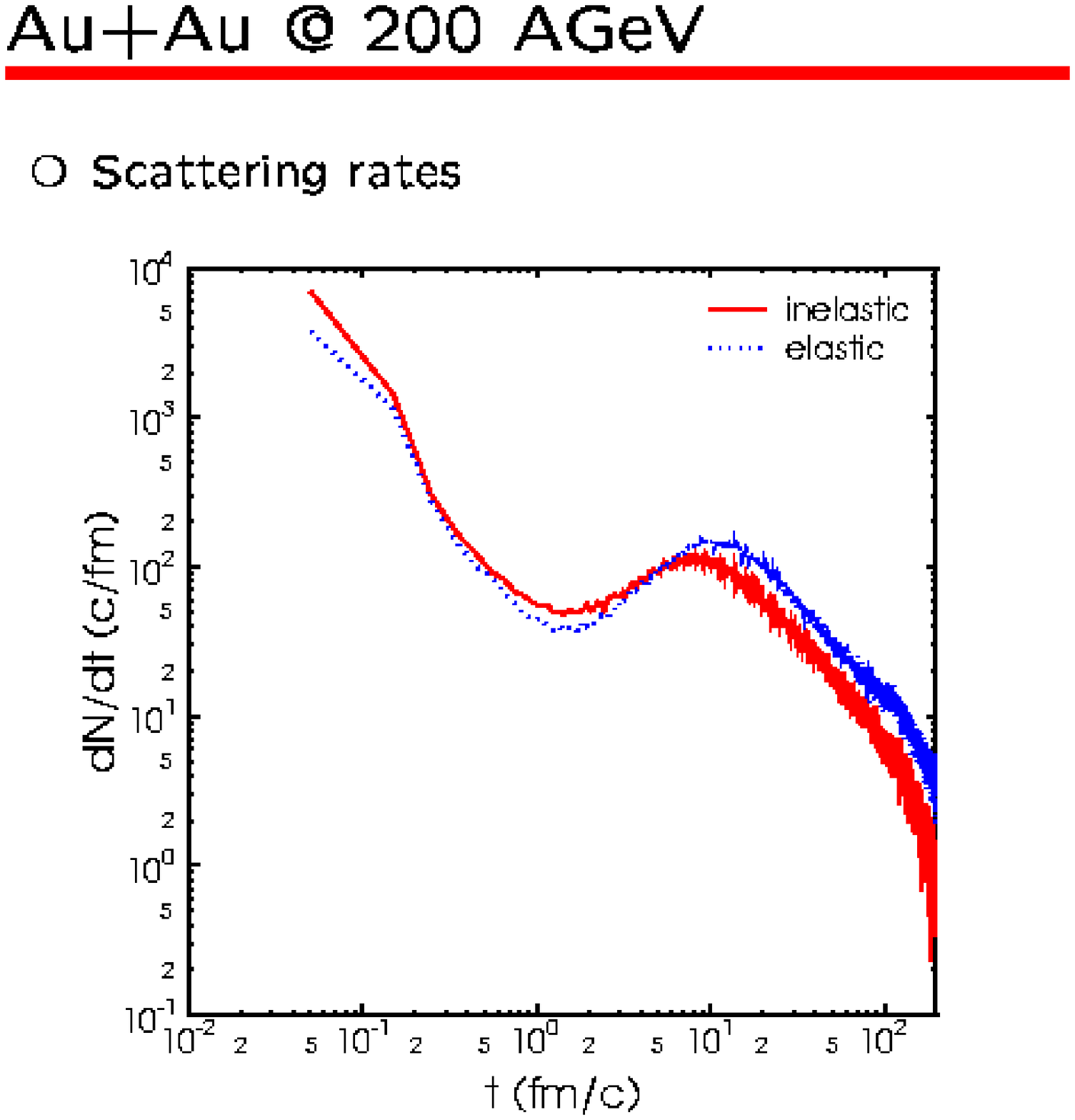}
\end{center}
\caption{\label{Fig10} Elastic and inelastic cross section averages
during the time evolution in a micro-dynamical simulation of a heavy
ion collision.  (From Marcus Bleicher's talk.)}
\end{figure}

\va
In order to understand better this interplay we need to gain information
on the process and on the achieved degree of equilibration. This is the
main goal of {microscopic computer simulation} codes,
reviewed and reported about by Sven Soff, Marcus Bleicher, Klaus Werner
and Christoph Hartnack. This field looks like a huge jungle, at least
for non-experts. It is not really transparent for us and looks like
the same green color everywhere. Efforts to incorporate QCD color
dynamics, if only on the semiclassical level, are rare.
We have listened for a while to the sounds from this jungle. A clear
tenor has emerged: all calculations conclude that simple string dynamics
would not do a good job in reproducing experimental data. Either
the string constant has to be changed or the hadronization mechanism
without deeper justification. A further negativum is the presence of
far too many internal code parameters, a positivum is the 
self-equilibrating property inherent in this approach. Fig.\ref{Fig10} 
shows elastic and inelastic rates, related to the maintenance of thermal
equilibrium and to the evolution of chemical composition, respectively.
A cross over of the respective dominance of these two rates is clearly
seen at times of the order of $10$ fm/c.

\va

\begin{figure}
\begin{center}
\includegraphics[width=0.5\textwidth,bb=30 -200 566 838]{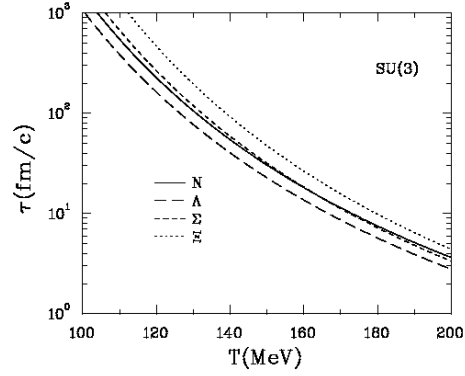}
\end{center}

\vspace{-50mm}
\caption{\label{Fig11} Characteristic production time of baryon -- 
antibaryon pairs from a thermal quark-gluon plasma.
(From Joe Kapusta's talk.}
\end{figure}

\va
There are two main problems microscopic codes suffer from:
\begin{enumerate}
\item	the initial state is unknown,
\item	the final state, in particular the dynamics of hadronization,
	is oversimplified.
\end{enumerate}
New progress has been presented by Joe Kapusta, who calculated thermal
baryon -- antibaryon production rates in a QGP by using effective
field theory (a quite realistic one). His result, stating a
characteristic time of about $10$ fm/c at the usually inferred temperature
gives some interesting perspective to improve microcodes in the
future (Fig.\ref{Fig11}).

\va

\begin{figure}
\begin{center}
\includegraphics[width=0.5\textwidth,bb=36 215 559 627]{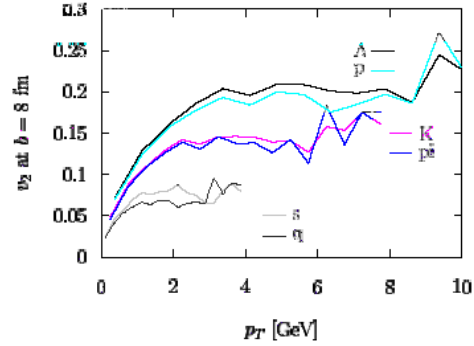}
\end{center}
\caption{\label{Fig12} Flow asymmetry of mesons, baryons and partons
using parton recombination in a numerical simulation. 
(From D\'enes Moln\'ar's talk.)}
\end{figure}

\va
Studying the dynamics of hadron formation is not simply a selfish
entertainment of pQCD adversed colleagues. D\'enes Moln\'ar has
pointed out that {parton recombination} may resolve the
so called opacity puzzle. Studying the flow asymmetry via the
Fourier expansion coefficient $v_2$ as a function of the transverse
momentum $p_T$ baryons, mesons and partons can well be separated
into three distinct curves (Fig.\ref{Fig12}). 
This {\em explains} why baryon
and meson (proton and pion) flows are different, and on the other
hand also reduces the need for assuming too many gluons in the
initial state used for the calculation. Very important data are
in this respect the flow asymmetry, $v_2$, the collectivity ratio
$R_{AA}$, and their flavor dependence. This is a ``h\'ap''
(pronounced: kha up) from the QGP duck -- at least this is what
Hungarian ducks say...

\va

\begin{figure}
\begin{center}
\includegraphics[width=0.5\textwidth,bb=36 208 559 633]{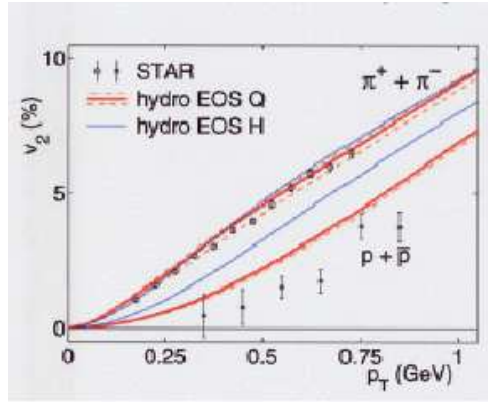}
\end{center}
\caption{\label{Fig13} Flow asymmetry for protons and pions in a    
hydrodynamical calculation using non-hadronic equation of state.
(From  Ulrich Heinz's talk.)}
\end{figure}

\va
Studying the very low $p_T$ flow asymmetry data in the framework
of {parton hydrodynamics} Ulrich Heinz made a remarkable
conclusion related to this issue\footnote{Since his talk was going
on by an express speed I hardly realized first what the matter is, 
but thank to talking him after his presentation he reinforced me in 
that, what I am going to say now.} The $v_2 - p_T$ data for protons and
pions from RHIC simultaneously {\em ``cannot be described by
hadronic equations of state''(EoS).} One has to use at least mixed
phase EoS to fit (Fig.\ref{Fig13}). The inferred initial state has an
enormously high initial energy density ($24$ GeV/fm$^3$) which lasts
at least $5$ fm/c in time. Another ``h\'ap'' from the duck.

\va

\va
These achievements bring me to mention upcoming efforts to
describe {hadronization dynamics} on the elementary level.
Ralf Rapp has given a review talk on this and many others, e.g.
Thomas Mehen, were presenting QCD based studies of heavy meson
formation. Here an expansion in terms of
$\alpha_sm_c\Lambda_{QCD}/m_T^2$ has been established.
However it is still unresolved that {\em what is the real process
of hadronization.} It is in particular unclear, what the
correct hadronic wave function in the final state is, and what is the
leading soft process. A clean-cut theory approach would here explain
the very fragmentation function applied and would recombine partons
to hadrons not just statistically, but by first-principle, non-perturbative
QCD dynamics.

\va
So we have arrived at {lattice QCD}. This is our only hope
in the moment to learn about these questions. Now we are dashing
in the water of Lattice Lake and are closer to the duck than ever before.
We just should proceed carefully, the QGP-duck may be a shy creature,
easy to whisk away.

\begin{figure}
\begin{center}
\includegraphics[width=0.7\textwidth,bb=38 401 817 772]{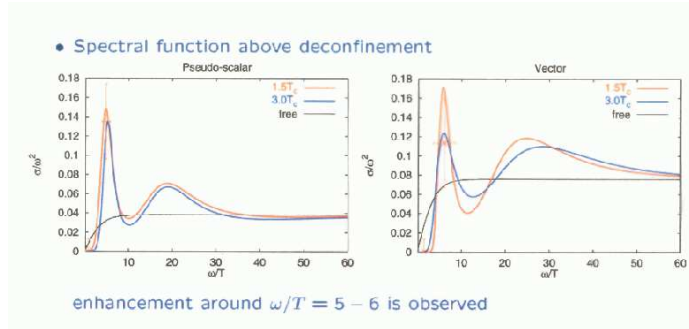}
\end{center}
\caption{\label{Fig15} Spectral functions from lattice QCD calculation
above the deconfinement temperature in the pseudoscalar and pseudovector
meson channel.  (From P\'eter Petreczky's talk.)}
\end{figure}

\begin{figure}
\begin{center}
\includegraphics[width=0.5\textwidth,bb=36 259 560 584]{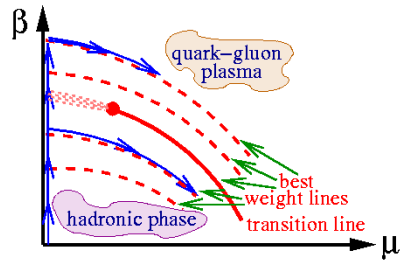}
\end{center}
\caption{\label{Fig16} A sketch of the re-weighting method used in finite
chemical potential lattice QCD calculations. (From Steven Gottlieb's talk.)}
\end{figure}

\va
The main {\bf new} lessons we have learned from lattice QCD at this
conference are as follows:
\begin{enumerate}
\item	The $q\overline{q}$ potential at temperatures moderately over
	$T_c$ does not resemble a picture of Debye screening
	(contrary to former expectations), but rather justifies a
	gradual string break picture.
\item	Studies of mesonic spectral functions reveal that up to
	$T=3T_c$ temperatures the deconfined phase is 
	{\em far from being a plasma of free quarks and gluons};
	plasmons alike pre-hadrons are formed. I would like to show here
	characteristic results presented by P\'eter Petreczky
	on Friday morning (cf. Fig.\ref{Fig15}).
\item	Finally, as we could have learned from the review talk of Steven
	Gottlieb, there are finite $\mu_B$ calculations by Zolt\'an
	Fodor and S\'andor Katz, and also other upcoming results at
	$\mu_B=0$ which all confirm that in the most realistic $2+1$
	flavor QCD with dynamical quarks {\em there is no first order
	phase transition at color deconfinement} just a crossover 
	(cf. Fig.\ref{Fig16}).
\end{enumerate}

\va

These important new results of lattice QCD should be taken seriously
also in our community. No naked quarks can be used for recombination
(only effective ones, and plasmons) and very probably there are less
gluons present at this process, as previously believed.
Nobody is entitled to take bag models as input for the phenomenological
calculation any more, rather the string picture should be
revitalized.

\va

\va
We do not need to be plainly passive in this field.
We should try to communicate our needs and wishes to lattice people.
What would we, what would I like to ask from them?
The main message for lattice Santa Claus has to be:
please, help to deal with hadron formation from the quark
gluon plasma. This wish can be expanded to a longer list:
\begin{itemize}
\item	Clarify the QGP structure near $T_c$. (Lattice QCD is already
	going into this direction).
\item	Carry out finite $\mu_B$ and not less finite $\mu_S$
	simulations. (The latter would be important also for
	strangelet and strange star search, for the main idea there
	is still built on the bag model. Now it is within reach
	that we can do better.)
\item	Give hints about the real time hadron formation.
	(Although this is usually considered to be {\em hopeless},
	it is {\em not impossible}. I shall elaborate on this
	remark a little bit later.)
\item	Inform about hadron properties near $T < T_c$ in the
	confining phase. (Mass, width, hadronic sizes are of interest.)
\item	Help to clarify string dynamics.
	(String melting to color rope and string breaking in case
	of non-trivial, higher color multiplets at the ends.)
\end{itemize}
Coming back to the question of squeezing real time information out
from lattice QCD, let me outline, what the problem blocking the way
here is. The main difference between real and imaginary time
treatment can be comprised into two versions of
the path integral formula. In the Euclidean version,
$$ \langle {\cal O} \rangle_{{\rm Euk}} \, = \,
\frac{1}{Z} \int {\cal D}U e^{-S[U]} {\cal O}(U), $$
${\cal O}(U)$ is the observable (any operator with physical meaning)
and ${\cal D}U exp(-S[U])$ is the weighted integral measure in 
the field-configuration space allowing for Monte Carlo type
simulation techniques to converge in an acceptable time to the
expectation values we are seeking for.

\va
The real (Minkowski) time version of this formula,
$$ \langle {\cal O} \rangle_{{\rm Min}} \, = \,
\frac{1}{Z} \int {\cal D}U e^{iS[U]} {\cal O}(U), $$
would lead us to treat $e^{iS[U]} {\cal O}(U)$ as an observable
(because the complex $exp(iS)$ factor is oscillating now and is therefore
useless for Monte Carlo). It would burden a hopeless task upon us:
to integrate over all possible configurations 
(with the measure ${\cal D}U$ only). On a lattice with ${\cal N}$
elements (sites, links, etc.) even in the case of a most simple-minded
two-state (Ising) model this means an order of $2^{{\cal N}}$
contributions to the wanted integral. This is exponentially
prohibitive, akin to algorithmic NP problems, which cannot be
solved in a time polynomial in ${\cal N}$.

\va
In principle there can be two ways out from this problem, one of them
may become viable already in the near future. At finite temperature
and short enough real-time problems the {re-weighting}
technique -- already applied in finite $\mu$ calculations -- can be
of use. The other way today still sounds like science fiction,
being hopeless yet not impossible: using quantum computers and
corresponding q-bit manipulating algorithms which would solve
the NP problem in general. We should watch the development in this
field in the future.

\begin{figure}
\begin{center}
\includegraphics[width=0.5\textwidth,bb=38 235 591 772]{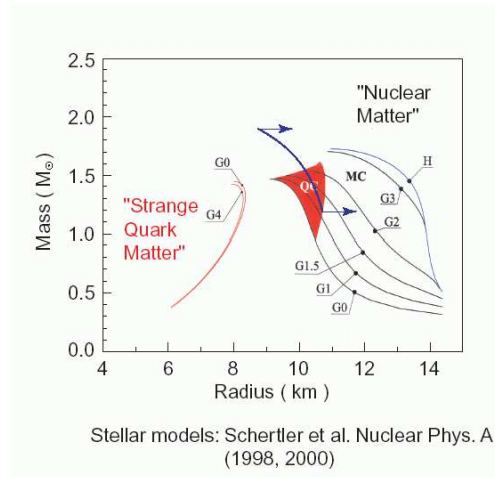}
\includegraphics[width=0.5\textwidth,bb=38 287 557 772]{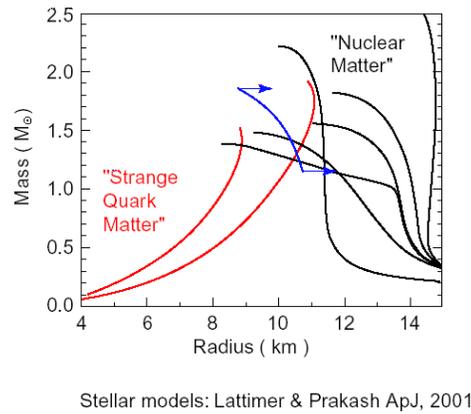}
\end{center}
\caption{\label{Fig17} Allowed regions of radius and mass of the   
RXJ1856 star compared with different calculations assuming strange
quark matter or other non-hadronic composition for a neutron star.
(From Markus Thoma's talk.)}
\end{figure}

\va
Last but not least our QGP-duck may be already flying. Flying in the
high skies of {astrophysics} and cosmology. 
Let me try to reconcile the main message here. The basic idea is
that if and whenever a star according to some observable
properties like size and brightness occurs to be abnormal,
two possible reasons could hide behind:
\begin{enumerate}
\item	the core or the whole of that (neutron) star consists of
	an exotic material with non-nuclear equation of state, or
\item	a misjudgment of either observable led us to a wrong conclusion.
\end{enumerate}
For such an exotic material we have heard quite a few suggestions.
Ariel Zhintinsky talked about Q-balls tightened together by axion
domain walls (unfortunately no sign of any axion has been observed yet),
Jeffrey Bowers about color crystals, Madappa Prakash about strange
quark matter, several others about color-flavor locking or kaon condensate.
Markus Thoma has shown us nice photos of the Chandra satellite
and of the star RXJ1856. This star was a candidate for being
extraordinarily compact with a radius of $R < 6$ km, and still too
bright in the X-ray spectrum. Several speakers concluded, however,
that in this case the size of this star has been misjudged.
The following two figures borrowed from Markus' talk show that
present astronomical data would exclude pure strange quark matter stars, 
but not hybrid stars with quark or mixed phase core (Fig.\ref{Fig17}).
Using another equation of state promoted by Prakash and Lattimer
there even could be a version of strange quark star compatible
with the allowed range.

\va

\va
Following this review of main themes and rhymes of this conference
I would now like to summarize the main achievements. My third
prepared transparency had two words on it: ``Explanations'' and
``Predictions''\footnote{I swear it was otherwise empty. Now I can
fill this form.}. Two nice, exclusively non-hadronic explanations of the
difference in proton and pion flow asymmetry at RHIC were given by
D\'enes Moln\'ar (based on parton recombination) and by Ulrich Heinz
(based on non-hadronic EoS in hydrodynamics).

\va
Also a few predictions has been made. Sean Gavin suggested a dcc
signal observable in charge fluctuations, Joe Kapusta predicted
thermal baryon - antibaryon pair creation rates from QGP.
An important prediction of lattice QCD was presented by P\'eter
Petreczky. He has shown that with $2\sigma$ confidence the
{\em $J/\psi$ ground state does not disappear above $T_c$}.
Even a quantitative value for the width of this resonance
has been given ($210$ MeV at $T=1.08T_c$).
Finally Prakash has made\footnote{according to himself just last week}
a prediction pointing out much smaller time scales for strange
quark stars in a neutron-star black hole binary system than for
normal neutron star partners, $\tau_{{\rm SQM}} \ll \tau_{{\rm NS}}$,
in the pre-merging stage. This should, in principle, be seen in the
modified spectra of gravitational waves.

\va
I am now almost at the end of this review, so -- as a summary of
the summary -- let me put on a list of dualisms. These are
dual views which have to be resolved in order to achieve further
progress in the theory of our field. In general dualisms can be
resolved in two ways: either by fusion, merging to a more complex
view\footnote{as  the particle -- wave duality
had been resolved by working out quantum mechanics},
or by decision, which view is correct.
\begin{itemize}
\item	{\em Thermal vs. combinatorial} models of hadronic
	final states. We hope to be able to decide here some day.
\item	{\em Parton vs. constituent quark} picture. Here an
	interpolating or unifying model should be searched for.
\item	{\em Real time vs. imaginary time} problems treated
	in lattice QCD, related to the question of statistical
	weights and observables. Here a case by case decision
	seems to be favorable.
\item	A very important dualism existed between 
	first order and higher order phase transition or 
	{\em crossover} at color deconfinement. This question
	now seems to approach clarification quickly.
\end{itemize}
Last but not least a shortlist of topics which were not on the
official program, although many of them have been mentioned or
referred to. Finite $\mu$ lattice calculations should have been worth
for a self-contained review. The quark recombination model ALCOR,
pioneered by J\'ozsef Zim\'anyi as well as ideas like colored
molecular dynamics, gluon saturation and colored glass condensate
were only indirectly talked about.

\va
Now we have arrived at the conclusion. The conclusion of the theory
part of the SQM2003 conference is straight and simple.
This conference has been a success because
\begin{itemize}
\item[a)] now we understand more than before, and
\item[b)] we have got some idea what needs to be done.
\end{itemize}
A side remark to the discussion about the future of this conference
series: please, do not change the name! We are already fond of it.
The phrase ``Strange Quark Matter'' does not exclude any new,
good physics ideas related to heavier flavors or other aspects.
Finally let me thank the organizers, first of all Berndt M\"uller
and Steffen Bass for providing us this inspiring atmosphere
and to you all for making the meeting to a success by your
contributions as speakers and listeners. Thank you!

\va \va
{\bf Acknowledgments} I thank the organizers for inviting me
to give the theoretical summary talk at 
this meeting and the Hungarian National Science Fund OTKA
for supporting my travel and participation,
as well as the research under contract Nr. T034269.

%%%%%%%%%%%%% FIGURES %%%%%%%%%%%
%\Figures

% \begin{figure}
% \begin{center}
% \includegraphics[width=0.5\textwidth,bb=49 228 546 614]{Fig14.ps}
% \end{center}
% \caption{\label{Fig14} Feynman graph for heavy meson formation.
%  (From Thomas Mehen's talk.)}
% \end{figure}

%%%%%%%%%%%%%%%%%% THE END %%%%%%%%%%%%%%
\end{document}